\documentclass[epj]{svjour}
\usepackage{graphics}
\usepackage{graphicx}
\usepackage{textcomp}
\usepackage{makeidx}
\usepackage{amsmath}
\usepackage{subfigure}
\usepackage{amssymb}
\usepackage{hyperref}
\usepackage{graphicx}
\usepackage{color}
\usepackage{float}
\usepackage{cite}
\usepackage{epstopdf}
\begin{document}
\title{Thermodynamic extended phase space and $P-V$ criticality of black holes at Pure Lovelock gravity }
\titlerunning{$P-V$ criticality of black holes at Pure Lovelock gravity }
\author{Milko Estrada\inst{1}  \thanks{\emph{e-mail:} milko.estrada@ua.cl} \and Rodrigo Aros \inst{2} \thanks{\emph{e-mail:} raros@unab.cl}}
\authorrunning{
M Estrada \and
R Aros
}            
\institute{Departamento de F\'isica, Facultad de ciencias b\'asicas, Universidad de Antofagasta, Casilla 170, Antofagasta, Chile. \and Departamento de Ciencias Fisicas, Universidad Andres Bello, Av. Republica 252, Santiago,Chile.  }

\date{Received: date / Revised version: date}
%
\abstract{ In this work the \textit{chemistry} of asymptotically AdS black hole, charged and uncharged, solutions of Pure Lovelock gravity is discussed. For this the mass parameter of black holes is identified with the enthalpy of the system together with the promotion of the cosmological constant to a thermodynamics variable proportional to the \textit{pressure} of the system. The equations of state for both, charged and uncharged, are obtained. It is shown that the charged case behaves as a Van der Waals fluid. The existence of a first order phase transition between small stable/large stable black hole, which is a reminiscent of the liquid/gas transition, is found. The critical exponents of the thermal evolution, for different cases of interest, are similar to those of the Van der Waals fluid.}

\PACS{
      {PACS-key}{discribing text of that key}   \and
      {PACS-key}{discribing text of that key}
     } 
%

\maketitle
\section{Introduction}

Certainly the existence of black holes was one of the most interesting predictions of General Relativity. However, the discovery that these objects, due to quantum fluctuations, emit as black bodies with temperatures dictated by the surface gravity \cite{Bekenstein:1972tm,Bekenstein:1973ur,Hawking:1974rv,Hawking:1974sw}, a purely geometric magnitude, simultaneously exposed  black holes as  thermodynamic objects and showed that black holes are scenarios where geometry and thermodynamics intertwine.

The first law of the black hole thermodynamics (see for instance Ref.\cite{Wald:1984rg}) is given by
\begin{equation}\label{1aleySinPdV}
dM=TdS+\Omega dJ+ \phi dQ,
\end{equation}
and represents the balance of energy through the modification of the macroscopic parameters of the black hole. Here $M$ corresponds to the mass parameter and was considered the internal energy of the system, namely the (ADM) mass. Finally, in Eq.(\ref{1aleySinPdV}) $T$ is the temperature computed as the $(4\pi)^{-1} \kappa$ with $\kappa$ is the surface gravity of the black hole horizon, $S$ is the entropy, $\Omega$ is the angular velocity, $J$ is the angular momentum, $\phi$ is the electrostatic potential and $Q$ is the electric charge.

By comparing Eq.(\ref{1aleySinPdV}) with the first law of thermodynamics one can notice the absence of the pressure/volume term, namely $-pdV$, which would stand for \textit{the macroscopic work} done by the system. In a matter of speaking, this was due to the lack of the concepts of volume and pressure for a black hole in the original derivation, based on accretion processes. Several works have studied this issue in the last 20 years \cite{Padmanabhan:2002sha,Tian:2010gn,Cvetic:2010jb,Dolan:2011xt}. To address this problem let us consider the first law of thermodynamics (see for example \cite{Dolan:2011xt}):
\begin{equation}\label{1aley1}
dU = TdS - pdV +\Omega dJ+ \phi dQ,
\end{equation}
where $U$ stands for the internal energy. To extend Eq.(\ref{1aleySinPdV}) to match Eq.(\ref{1aley1}), one can think of assigning a volume to the black hole by considering the \textit{volume}  defined its radius, for example $V=\frac{4}{3}\pi r_+^3$ for $d=4$ case. Unfortunately, since entropy is a function of the horizon radius as well, then Eq.(\ref{1aley1}) would be inconsistent due to $dS$ and $dV$ would not be independent directions. To address this problem, in Ref.\cite{Kastor1}, was proposed to reinterprete the mass parameter as the Enthalpy of the black hole, instead of internal energy $U$. Moreover, the cosmological constant was connected with the thermodynamic pressure. The law obtained is called the first law of (black hole) thermodynamics in the {\it extended phase space}. In Ref.\cite{MotorTermico1}, on the other hand, the promotion of the mass parameter to the enthalpy is based on the fact that to form a black hole would require to cut off a region of the space, and therefore an initial energy equal to $E_0 = -\rho V$, with $\rho$ the energy density of the system, is needed. In four dimensions the thermodynamics volume corresponds to $V=\frac{4}{3}\pi r_+^3$. Moreover, the presence of the cosmological constant defines $\rho = -P$ and therefore the mass parameter can be considered equivalent to
\begin{equation}
M=U-\rho V = U+pV,
\end{equation}
Here $M$ is to be recognized as the enthalpy $H$ of the system. With this in mind, the first law, in this  extended phase space, yields
\begin{equation} \label{1aley2}
dM=dH=TdS+Vdp+\phi dQ+\Omega dJ.
\end{equation}
This extended first law (\ref{1aley2}) also was derived in reference \cite{Mann1} by using Hamiltonian formalism. It is worth to mention that the definition of the extended phase space have allowed to construct a {\it heat engine} in terms of a black hole, see some examples in references \cite{MotorTermico1,MotorTermico5,MotorTermico7,MotorTermico13,MotorTermico14,MotorTermico15}, adding a new layer to our understanding of the black hole thermodynamics. Another interesting applications is the Joule Thompson expansion for black holes studied in \cite{JT1,JT2,JT3,JT4}.

\subsection{Phase Transitions}
The study of phase transitions in black hole physics has called a renewed attention in the last years due to the AdS/CFT conjecture. For instance, it is well known that the {\it Hawking-Page phase transition} \cite{Hawking:1982dh}, in the context of the AdS/CFT correspondence, has been re-interpreted as the plasma gluon confinement/deconfinement phase transition in the would-be dual (conformal) field theory. Similarly, the AdS Reissner Nordstr\"om's transitions in the $(\phi-q)$ diagram have been interpreted as liquid/gas phase transitions of Van der Waals fluids \cite{Chamblin:1999tk,Chamblin:1999hg}.

Recently the analysis of the $p-V$ critical behaviors (in the extended phase space) have been under studied extensively\cite{RegularBHquimica, PV1,PV2,PV3,PV4,PV5,PV6,PV7,PV8,PV9,PV10,PV11,PV12,Hendi:2017fxp,Hendi:2015cqz,Hendi:2015hoa,Hendi:2014kha}. For instance, in \cite{Mann2} was studied in the context of the charged $4D$ AdS black holes how the phase transitions between small/large black hole are analogous to liquid/gas transitions in a Van der Waals fluid. Moreover, it was also shown that the critical exponents, near the critical points, recovers those of Van der Waals fluid with the same \textit{compressibility factor} $Z=3/8$. In reference \cite{Mann1} was introduced a \textit{new interpretation} of the Hawking Page phase transition \cite{Hawking:1982dh} mentioned above, but in the context of $p-V$ critical behavior. 

\subsection{Higher Dimensions, Lovelock and thermodynamics}

During the last years 50 years several branches of theoretical physics have noticed that considering higher dimensions is plausible. Now, considering higher dimension in gravity opens up a range of new possibilities that retain the core of the Einstein gravity in four dimensions. Lovelock is a one of these possibilities as, although includes higher powers of curvature corrections, its equations of motion are of second order and thus causality is still insured. Generic Lovelock theory is the sum of the Euler densities $L_n$ in lower dimensions ($2n \leq d$) multiplied by coupling constants $\alpha_n$ \footnote{In even dimensions the maximum order, $n=d/2$, is in turn the corresponding Euler density of the dimension and therefore does not contribute to the equations of motion,}. The Lagrangian is
\begin{equation}
 \displaystyle L  =  \sum_{n=0}^{[d/2]} \alpha_n L_n,
\end{equation}
where $L_n=\frac{1}{2^{n}}\,\sqrt{-g}\ \delta_{\nu_{1}...\nu_{2n}}^{\mu_{1}...\mu_{2n}}\ R_{\mu_{1}\mu_{2}}^{\nu_{1}\nu_{2}}\cdots R_{\mu_{2n-1}\mu_{2n}}^{\nu_{2n-1}\nu_{2n}}$. The $L _0 \propto 1$ term is related by the cosmological constant, $L_1 \propto R $ is related by the Ricci scalar, and $L_2 \propto R^{\alpha\beta}_{\hspace{2ex}\mu\nu}R^{\mu\nu}_{\hspace{2ex}\alpha\beta}-4 R^{\alpha\nu}_{\hspace{2ex}\beta\nu} R^{\beta\mu}_{\hspace{2ex}\alpha\mu} + R^{\alpha\beta}_{\hspace{2ex}\alpha\beta}R^{\mu\nu}_{\hspace{2ex}\mu\nu}$ is the Gauss Bonnet density. The EOM are:
\begin{align}\label{LovelockGenerico}
\frac{1}{\sqrt{g}}\frac{\delta }{\delta g_{\mu\nu}} (L\sqrt{g}) &= G^{\mu}_{ \hspace{1ex} \nu} \nonumber \\
&= \sum_{n} \alpha_{n} \frac{1}{2^n} \delta^{\mu_1 \ldots \mu_{2n} \mu}_{\nu_1 \ldots \nu_{2n} \nu} R^{\nu_{1} \nu_{2}}_{\hspace{2ex}\mu_{1} \mu_{2}}\ldots R^{\nu_{2n-1} \nu_{2n}}_{\hspace{2ex}\mu_{2n-1} \mu_{2n}}
\end{align}
where $G^{\mu \nu}_{(LL)}$   which, by definition, satisfies $\nabla_\mu G^{\mu\nu}_{(LL)} \equiv 0$.

In higher curvature the $p-V$ criticality have been studied for example in reference \cite{Mann3} for generic Lovelock black holes with conformal scalar hair arising from coupling of a real scalar field to the dimensionally extended Euler densities and in reference \cite{Nam:2019clw} for a regular charged AdS black hole in Einstein Gauss Bonnet gravity. See also \cite{Hendi:2015soe,Xu:2014tja}.

One drawback of a generic Lovelock gravity is the existence of more than single ground state, namely more than a single constant curvature spaces solution, or equivalently more than a single potential \textit{effective cosmological constants} \cite{Camanho:2011rj}. However, there are two families of Lovelock gravities that have indeed a single ground state. The first family has been originally studied in \cite{Banados:1994ur} and has a unique but $k$ degenerated ground state. 

The second case is usually called {\it Pure Lovelock gravity}. In this case the Lagrangian is a just the $n$-single term of Lagrangian plus a cosmological constant, \textit{i.e.}, $L = \alpha_n L_n + \alpha_0 L_0$. This has a single and non-degenerated ground state of negative curvature $-1/l^2$ for $n$ odd. Several solutions of Pure Lovelock theories are known. For instance vacuum black hole solutions can be found in references \cite{Cai:2006pq,DadhichBH,Dadhich3,PureLovelock1,PureLovelock2,Aranguiz:2015voa} and regular black holes in references \cite{milko1,milko2}. See other applications in references \cite{Dadhich1,Dadhich2,Dadhich4,Dadhich5}.

In the next sections the thermodynamic of the Pure Lovelock solutions will be carried out in details. For this, the cosmological constant will be promoted to the intensive thermal pressure of the system. Under this assumption, it will be shown that the system behaves as Van der Waals fluid. Finally, the critical coefficient near the critical points will be computed.

\subsubsection{{\bf Vacuum black hole in Pure Lovelock gravity: Uncharged Asymptotically AdS Case}}

Let us consider a static geometry, represented in Schwarzschild coordinates by
\begin{equation}\label{linelement}
ds^2 =-f(r) dt^2+ \frac{dr^2}{f(r)} + r^2 d \Omega^2_{d-2}.
\end{equation}
For simplicity the cosmological constant is fixed such that
\begin{equation} \label{Lambda1}
    \Lambda = - \frac{(d-1)(d-2)}{2 l^{2n}}.
\end{equation}
This is similar to the definition in \cite{Aranguiz:2015voa} but a numerical factor. The equations of motion reduce to merely
\begin{equation}
\frac{d}{dr} \left ( r^{d-2n-1}\big (1-f(r) \big)^n +  \dfrac{ r^{d-1}}{{l}^{2n}} \right ) =0.
\end{equation}
Solutions of these equations have been studied in references \cite{Cai:2006pq,Dadhich1}. In terms of Eq.(\ref{linelement}) the solution is defined by
\begin{equation} \label{funcionPLvacio}
f(r) =1 - \left (\frac{2M}{r^{d-2n-1}} - \frac{r^{2n}}{l^{2n}} \right )^{1/n} .
\end{equation}
One can notice that $f(r)$ can take complex values for even $n$. Because of this, in this work will be considered only odd $n$. The thermodynamic pressure can be read off this definition as
\begin{equation} \label{p1}
    p=-\frac{\Lambda}{(d-2)\Omega_{d-2}},
\end{equation}
where $\Omega_{d-2}$ is the unitary area of a $d-2$ sphere. It is worth to stress that Eqs.(\ref{Lambda1},\ref{p1}) coincide with the definitions in Ref.\cite{Mann1,Mann2} for $n=1$ and $d=4$. Now, replacing Eqs.(\ref{Lambda1},\ref{p1}) into Eq.(\ref{funcionPLvacio}) yields
\begin{equation} \label{funcionPLvacio1}
    f(r)=1 - \left (\frac{2M}{r^{d-2n-1}} - \frac{2 \Omega_{d-2}r^{2n}}{d-1}p  \right )^{1/n},
\end{equation}
where the dependence of $f(r)$ on $p,M$ have been made explicit.

\subsubsection{{\bf Charged Pure Lovelock solution}}
The charged case is slightly difference since it is necessary to solve the Maxwell equations and to include an energy momentum tensor into the gravitational equations. Let us start by defining $A_\mu = A_t(r) \delta_\mu^t$, which defines only the non-vanishing component of the Maxwell tensor 
\begin{equation} \label{PotencialVectorial}
    F_{tr}=-\partial_rA_t(r), 
\end{equation}
and the single non vanishing component of the Maxwell equations, 
\begin{equation} \label{TensorElectromagnetico}
  \nabla_{\mu}F^{\mu\nu} =0 \rightarrow  \frac{d}{dr} \left ( r^{d-2} F_{tr} \right) =0.
\end{equation}
Likewise, the gravitational equations are reduced to the single independent component
\begin{equation}\label{GravitationalEMEq}
 \frac{d}{dr} \left ( r^{d-2n-1}\big (1-f(r) \big)^n +  \dfrac{ r^{d-1}}{{l}^{2n}} \right ) = (F_{tr})^2 r^{d-2}.
\end{equation}
It is straightforward to integrate Eqs.(\ref{TensorElectromagnetico},\ref{GravitationalEMEq}) yielding \cite{Aranguiz:2015voa}
\begin{equation} \label{funcionPLvacioQ}
f(r) =1 - \left (\frac{2M}{r^{d-2n-1}} - \frac{r^{2n}}{l^{2n}} - \frac{Q^2}{(d-3) r^{2d-2n-4}} \right )^{1/n},
\end{equation}
and
\begin{equation}
    F_{tr}=\frac{Q}{r^{d-2}} \rightarrow A_t(r)= \phi_\infty + \frac{Q}{(d-3)r^{d-3}},
\end{equation}
As usual $\phi_{\infty}=0$ is fixed such that $\displaystyle \lim_{r\rightarrow \infty} A_t(r) =0$. 

Now,  by direct observation, one can notice that for even $n$ the existence of certain ranges of $r$ where $f(r)$ can take complex values. To avoid this only odd $n$ will be considered from now on.

As done previously, by replacing Eqs(\ref{Lambda1},\ref{p1}) into Eq.(\ref{funcionPLvacioQ}), $f(r)$ can be written in terms of the thermodynamics variables as
\begin{equation} \label{funcionPLvacio1Q}
    f(r)=1 - \left (\frac{2M}{r^{d-2n-1}} - \frac{2 \Omega_{d-2}r^{2n}}{d-1}p  - \frac{Q^2}{(d-3) r^{2d-2n-4}}   \right )^{1/n}.
\end{equation}

\section{Extended phase space in Vacuum Pure Lovelock Gravity}

In this section the extended phase space formalism, mentioned above, will be introduced to analyze the solutions. One can notice that the fist law of the thermodynamics, Eq.(\ref{1aley2}), for the non rotating case takes the form
\begin{equation} \label{1aleyconcarga}
   dH= dM=TdS+Vdp + \phi dQ.
\end{equation}
Now, in order to construct a thermodynamic interpretation one must notice that under any transformation of the parameters the function $f(r_+,M,Q,l)$  must still vanishes, otherwise the transformation would not be mapping black holes into black holes in the space of solutions. Indeed, $\delta f(r_+,M,Q,l) = 0$ and $f(r_+,M,Q,l)=0$ are to be understood as constraints on the evolution along the space of parameters. However, there is another approach by recalling that the mass parameter, $M$, is also to be understood as a function of the parameters $M(r_+,l,Q)$ as well.

Since the thermodynamic parameters are $S,p$ and $Q$, therefore it is convenient to reshape $M=M(S,p,Q)$ in order to explicitly obtain 
\begin{equation} \label{1aley3}
    dM= \left ( \frac{\partial M}{\partial S}    \right)_{p,Q}dS + \left ( \frac{\partial M}{\partial p}    \right)_{S,Q}dp + \left ( \frac{\partial M}{\partial Q}    \right)_{p,S}dQ .
\end{equation}
This corresponds to the definitions of the component of the tangent vector in the space of parameters, but also correspond to the definitions of the temperature, thermodynamic volume and electric potential in the form of
\begin{eqnarray}
T &=&\left ( \frac{\partial M}{\partial S}    \right)_{p,Q},\label{temperatura} \\
V &=& \left ( \frac{\partial M}{\partial p}    \right)_{S,Q} \textrm{ and }\label{volumen} \\
\phi &=& \left ( \frac{\partial M}{\partial Q}    \right)_{S,p}. \label{potencial}
\end{eqnarray}

On the other hand, the variation along the space of parameters of the condition defined by $f(r_+,M,p,Q)=0$,
\begin{eqnarray}
  df(r_+,M,p,Q) &=& 0\label{variacion}\\
   &=& \frac{\partial f}{\partial r_+}  dr_+ + \frac{\partial f}{\partial M} dM + \frac{\partial f}{\partial p} dp + \frac{\partial f}{\partial Q} dQ,\nonumber
\end{eqnarray}
yields a second expression for $dM$ given by
\begin{align} \label{variacion1}
    dM &=  \left (\frac{1}{4\pi} \frac{\partial f}{\partial r_+} \right ) \left( -\frac{1}{4 \pi} \frac{\partial f}{\partial M} \right)^{-1} dr_+ \nonumber \\
    &+ \left( - \frac{\partial f}{\partial M} \right)^{-1} \left ( \frac{\partial f}{\partial p} \right ) dp +\left( - \frac{\partial f}{\partial M} \right)^{-1} \left ( \frac{\partial f}{\partial Q} \right ) dQ,
\end{align}
which must coincide with equation (\ref{1aley3}). In Eq.(\ref{variacion1}) one can recognize presence of the temperature, which geometrically is defined as
\begin{equation} \label{temperaturaconocida}
    T= \frac{1}{4\pi} \frac{\partial f}{\partial r_+},
\end{equation}
 yielding 
\begin{equation} \label{dS1}
\left( -\frac{1}{4 \pi} \frac{\partial f}{\partial M} \right)^{-1} dr_+= dS,
\end{equation}
which is a very known result. Now, by the same token, the thermodynamic volume and electric potential are given by
\begin{equation} \label{volumen1}
    V=\left ( \frac{\partial M}{\partial p}    \right)_{S,Q}= \left( - \frac{\partial f}{\partial M} \right)^{-1} \left ( \frac{\partial f}{\partial p} \right ) 
\end{equation}
and
\begin{equation} \label{potencial1}
    \phi = \left ( \frac{\partial M}{\partial Q}    \right)_{S,p}= \left( - \frac{\partial f}{\partial M} \right)^{-1} \left ( \frac{\partial f}{\partial Q} \right ),
\end{equation}
respectively. These expressions will be discussed in the next sections.

\subsection{Smarr Expression}
Considering Plank units, one can notice that $p$, the pressure, has units of $[p]=\ell^{-2n}$. See Eqs.(\ref{Lambda1},\ref{p1}, \ref{presionPL1}). Likewise, one can check that $[M] = \ell^{d-2n-1}$, $[Q] = \ell^{d-n-2}$ and $[S] = \ell^{d-2n}$. 

Following {\it Euler's theorem} \cite{Altamirano:2014tva}, with $M(S,p,Q)$, one can construct the {\it Smarr formula} for Pure Lovelock gravity given by
\begin{equation}
    \frac{d-2n-1}{d-2n}M=TS+\frac{d-n-2}{d-2n} \phi Q - \frac{2n}{d-2n}VP.
\end{equation}
This coincides with the definitions discussed in \cite{Altamirano:2014tva,Mann2,Mann1} for $n=1$ . Derivations of the Smarr formula for generic Lovelock theory are discussed in \cite{Dolan:2014vba,Kastor:2010gq}. 

\subsection{ Uncharged asymptotically AdS}
Replacing the solution of Eq.(\ref{funcionPLvacio1}) into Eq.(\ref{dS1}) yields
\begin{align}
\left( -\frac{1}{4 \pi} \frac{\partial f}{\partial M} \right)^{-1} dr_+&=2\pi n r_+^{d-2n-1} \nonumber \\
&=d \left ( \frac{2}{d-2n}n \pi r_+^{d-2n}  \right ) = dS,
\end{align}
or equivalently
\begin{equation} \label{entropia1}
    S= \frac{2}{d-2n}n \pi r_+^{d-2n},
\end{equation}
which coincides with reference \cite{Cai:2006pq}. Eq.(\ref{volumen1}), on the other hand,  yields
\begin{equation} \label{volumen11}
    \left( - \frac{\partial f}{\partial M} \right)^{-1} \left ( \frac{\partial f}{\partial p} \right )  = V = \frac{\Omega_{d-2}}{d-1}r_+^{d-1},
\end{equation}
which corresponds to the volume of a $(d-2)$ sphere of radius $r_+$ . This coincides with the definition in Refs.\cite{Mann1,Mann2}.

\subsubsection{{ \bf Fluid equation of state}}
It is worth to notice at this point that the temperature, see Ref.\cite{milko1}, can be expressed as 
\begin{equation} \label{TemperaturaPL1}
    4\pi n T= \frac{d-2n-1}{r_+} + (d-1)\frac{ r_+^{2n-1}}{l^{2n}},
\end{equation}
where, the temperature has units of $\ell^{-1}$. One can make explicit the dependence on the pressure, by inserting Eqs.(\ref{Lambda1},\ref{p1}) into Eq.(\ref{TemperaturaPL1}). Therefore, 
\begin{equation} \label{presionPL1}
    2 \Omega_{d-2}p = 4 \pi n \frac{T}{r_+^{2n-1}} - \frac{d-2n-1}{r_+^{2n}}.
\end{equation}

\subsubsection{\bf Physical Pressure}
Before to proceed a digression is necessary. As mentioned above $[p] = \ell^{-2n}$, however the physical pressure, $p_G$, must be satisfied [Force/Area]$=\ell_p^{-2}$, since the area has units of [area]$=\ell_p^{d-2}$. In the literature $p_g$ is called the geometrical pressure. Since $p$ and $p_G$ must be connected by $p_G \sim \alpha_n p$, namely $[p_{G}] = \left [\alpha_n p \right]=\ell_p^{-d}$, therefore
 \begin{itemize}
    \item the {\it coupling constants} must satisfy $[\alpha_n]=\ell^{2n-d}$ \cite{Oliva1}. For $n=1$ this coincides with the inverse of the higher dimensional Newton constant $G_d^{-1}$ \cite{Maartens:2003tw} and
    \item  there is still room for a dimensionless constant which can be used to adjust the definition. 
\end{itemize}
With this in mind,  $p_G$ can be taken as 
\begin{equation}
    p_{G} =2 \Omega_{d-2} \ell_p^{2n-d} p = 4 \pi n \frac{T}{\ell_p^{d-2n}r_+^{2n-1}} + \ldots,
\end{equation}
which, in turns, defines the \textit{specific volume} of the system as
\begin{equation} \label{volumenespecificoPL}
    v=\frac{\Omega_{d-2}}{2\pi} \ell_p^{d-2n}r_+^{2n-1}.
\end{equation}

Notice that $v$ indeed has units of volume, namely $[v]= \ell_p^{d-1}$ \cite{Hennigar:2018cnh}. Finally, after some replacements, 
\begin{equation} \label{presionPL2}
   p = \frac{n T}{v} - \frac{1}{2\Omega_{d-2}(2\pi/\Omega_{d-2})^{2n/(2n-1)}}\frac{d-2n-1}{v^{2n/(2n-1)}},
\end{equation}
which can be recognized as the Van der Walls equation for $n=1$, namely $P=T/(v-b)-a/v^2$, with $b=0$ \cite{Mann1}.  

\subsubsection{{\bf Hawking-Page phase transition}}

In reference \cite{Hawking:1982dh} was analyzed the thermodynamic behavior of Schwarzschild AdS space, which differs from the Schwarzschild case due to the presence of the AdS gravitational potential,  namely the presence of $\sim r^2/l^2$ in $f(r)$ \cite{Czinner:2015eyk}. In this case there is a phase transition between black hole and AdS radiation at a critical temperature $T_{HP}$ where the Gibbs free energy, $G=M-TS$ vanishes \cite{Mann2,Czinner:2015eyk,Wang:2019vgz}. 

\begin{figure}
\centering
{\includegraphics[width=3in]{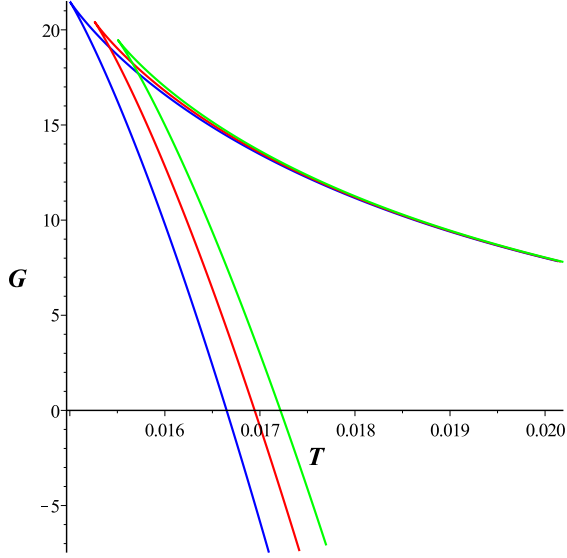}}
\caption{HP phase transitions for $n=3$, $d=10$ and $Q=1$, $p=0,0000045$(blue), $p=0,0000050$(red), $p=0,0000055$ (green). T (horizontal axis) v/s G (vertical axis). }
\label{HP}
\end{figure}

Fig.\ref{HP} displays the numerical behavior of the Gibbs free energy v/s temperature for $n=3$ and $d=10$ for Pure Lovelock. The upper curve represents the unstable small black hole (namely with negative heat capacity) and the lower curve represents the stable large black hole. It is direct to show that this behavior is similar for any other set of values of $n$ and $d$. 

\begin{figure}
\centering
{\includegraphics[width=3in]{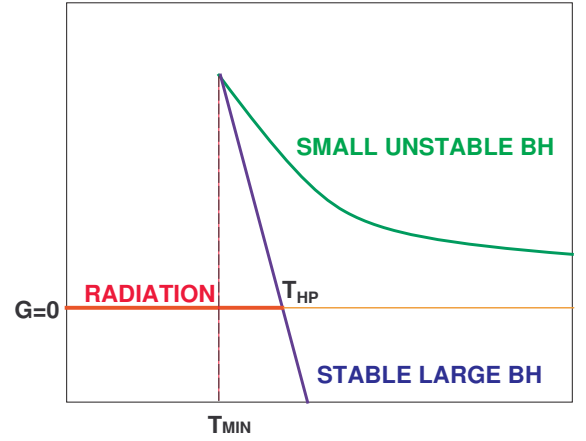}}
\caption{Generic behavior of Gibbs free energy (vertical axis) v/s T (horizontal axis). HP phase transition at $T_{HP}$}
\label{HPgenerico}
\end{figure}

The generic behavior is displayed in Fig.\ref{HPgenerico}. The upper green curve represents the unstable small black hole, the lower blue curve represents the stable large black hole, and the orange curve represents the thermal AdS radiation whose Gibbs free energy vanishes \cite{Mann1,Czinner:2015eyk,Wang:2019vgz,Costa:2015gol}. One can notice that for $[T_{min},T_{HP}[$ the large stable black hole has positive Gibbs free energy,  therefore, the preferred state corresponds to the thermal AdS radiation. On the other hand, for $T>T_{HP}$, the preferred state is the large stable black hole whose Gibbs free energy is negative. This hints the existence of a Hawking Page phase transition between radiation and the large black hole states at $T=T_{HP}$.  

Finally, it is worth to notice, from Fig.\ref{HP}, that the value of $T_{HP}$ increases as the pressure increases. This behavior is similar to the HP phase transition for the Schwarzschild AdS black holes \cite{Mann1,Belhaj:2015hha} or the polarized AdS black holes \cite{Costa:2015gol}.

\subsection{ Charged Pure Lovelock solution}
By replacing Eq.(\ref{funcionPLvacio1Q}) into Eqs.(\ref{dS1},\ref{volumen1}) the expressions for the entropy (\ref{entropia1}) and the volume (\ref{volumen11}) are obtained . On the other hand, replacing Eq.(\ref{funcionPLvacio1Q}) into Eq.(\ref{potencial1}) yields 
\begin{equation} \label{potencial2}
    \phi= \frac{Q}{(d-3)r^{d-3}}.
\end{equation}
This thermodynamics quantities, obtained by mean of variation of parameters, are consistent with the values presented in the literature. 

\subsubsection{\bf Fluid equation of state}
In this case the temperature can be written as
\begin{equation} \label{TemperaturaPL1Q}
    4\pi n T= \frac{d-2n-1}{r_+} + (d-1)\frac{ r_+^{2n-1}}{l^{2n}}- \frac{Q^2}{r_+^{2d-2n-3}} .
\end{equation}
By inserting equations (\ref{Lambda1},\ref{p1}) into equation (\ref{TemperaturaPL1Q}) is obtained
\begin{equation} \label{presionPL2Q}
    2 \Omega_{d-2}p = 4 \pi n \frac{T}{r_+^{2n-1}} - \frac{d-2n-1}{r_+^{2n}} + \frac{Q^2}{r_+^{2d-4}}.
\end{equation}
Now, by using the definition of the specific volume, defined in Eq.(\ref{volumenespecificoPL}) with $\ell_p=1$,
\begin{eqnarray}
 p &=& \frac{n T}{v} - \frac{1}{2\Omega_{d-2}(2\pi/\Omega_{d-2})^{2n/(2n-1)}}\frac{d-2n-1}{v^{2n/(2n-1)}} \nonumber \\
   &+& \frac{1}{2\Omega_{d-2}(2\pi/\Omega_{d-2})^{(2d-4)/(2n-1)}}\frac{Q^2}{v^{(2d-4)/(2n-1)}}. \label{presionPL3}
\end{eqnarray}
One can notice the similarity of this relation with the corresponding one for a Van der Waals fluid for $n=1$ but with the presence of an addition third term.  

\subsubsection{\bf Critical points and compressibility factor}
To compare Eq.(\ref{presionPL3}) with the behavior of a Van der Waals fluid it is necessary to determine the critical points of the system. The second order critical points are defined by the conditions,  
\begin{equation}
    \frac{\partial p}{\partial v}=0 \textrm{ and } \frac{\partial^2 p}{\partial v^2}=0.
\end{equation}
These determine the critical values
\begin{equation}
    v_c = \frac{\Omega_{d-2}}{2\pi}\left ( \frac{2d^2-2dn-7d+4n+6}{n(d-2n-1)}Q^2   \right )^{(2n-1)/(2d-2n-4)} ,
\end{equation}
and
\begin{align}
    T_c &= \frac{n(d-2n-1)}{2\pi n (2n-1)(2\pi/\Omega_{d-2})^{1/(2n-1)}v^{1/(2n-1)}} \nonumber \\
        &-\frac{(d-2)Q^2(2\pi/\Omega_{d-2})^{(2n-2d+4)/(2n-1)}v^{(2n-2d+4)/(2n-1)}}{2\pi n (2n-1)(2\pi/\Omega_{d-2})^{1/(2n-1)}v^{1/(2n-1)}}
\end{align}
Notice that $p_c=p(v_c,T_c)$ can be determined from equation (\ref{presionPL3}) by evaluation on the critical values $T_c$ and $v_c$. 

In the table (\ref{tabla1}) critical values $v_c$, $T_c$ and  $p_c$ and the {\it compressibility factor} $Z=p_cv_v/T_c$ are displayed for different values of $n$ and $d$ . For $n=1$ and $d=4$ the compressibility factor has the exact value $Z=3/8$ which coincides with the value of the compressibility for a Van der Waals fluid. In general, compressibility factor can be written as
\begin{equation} \label{factorcompresibilidad}
    Z=\frac{2d-2n-3}{4(d-2)},
\end{equation}
which implies that $Z<1$ extrictly. This implies that this can be interpreted as a low pressure gas. 
\begin{table*} 
\caption{Critical values and Compressibility Factor ($Z$) .}
\begin{tabular*}{\textwidth}{@{\extracolsep{\fill}}lccccc@{}}
\hline
\multicolumn{1}{c}{$n$   } & \multicolumn{1}{c}{$d$ } & \multicolumn{1}{c}{$v_c$}& \multicolumn{1}{c}{$T_c$}&\multicolumn{1}{c}{$p_c$}&\multicolumn{1}{c}{$Z=\dfrac{p_cv_c}{T_c}$} \\
 \hline
$1$ & $4$  & $\Omega_{d-2}/(2\pi)Q\sqrt{6}$ & $1/(18\pi Q) \sqrt{6}$ &$1/(24\Omega_{d-2}Q^2)$ & $3/8$   \\
\hline 
$1$ & $5$  & $\Omega_{d-2}/(4\pi)(120Q^2)^{1/4}$ & $4/(75\pi)(6750/Q^2)^{1/4}$ &$2/(45\Omega_{d-2}Q)\sqrt{30}$ & $5/12$   \\
\hline
$1$ & $6$  & $\Omega_{d-2}/(6\pi)(6804Q^2)^{1/6}$ & $9/(98\pi)(201.684/Q^2)^{1/6}$ &$9/(112\Omega_{d-2})(294/Q^2)^{1/3}$ & $7/16$   \\
\hline
$3$ & $8$  & $\Omega_{d-2}/(2\pi)(14Q^2)^{5/6}$ & $3/(490\pi) (14^5/Q^2)^{1/6}$ &$1/(280\Omega_{d-2}Q^2)$ & $7/24$   \\
\hline
$3$ & $9$  & $\Omega_{d-2}/(4\pi)(21^5\cdot 2^3Q^{10})^{1/8}$ & $8/(945\pi) (21^7\cdot 2/Q^2)^{1/8}$ &$4/(735\Omega_{d-2})(21\cdot 2^3/Q^6)^{1/4}$ & $9/28$   \\
\hline
$3$ & $10$  & $\Omega_{d-2}/(18\pi)(792)^{1/2}Q$ & $3/(242\pi) (2^7\cdot 11^9 \cdot 9/Q^2)^{1/10}$ &$9/(704\Omega_{d-2})(726/Q^6)^{1/5}$ & $11/32$   \\
\hline
$5$ & $12$  & $\Omega_{d-2}/(2\pi)(22Q^2)^{9/10}$ & $5/(2178\pi)(22^9/Q^2)^{1/10}$ &$1/(792\Omega_{d-2}Q^2)$ & $11/40$   \\
\hline
\end{tabular*} \label{tabla1}
\end{table*}

\subsubsection{\bf $P-v$ curve.}

\begin{figure}
\centering
{\includegraphics[width=3in]{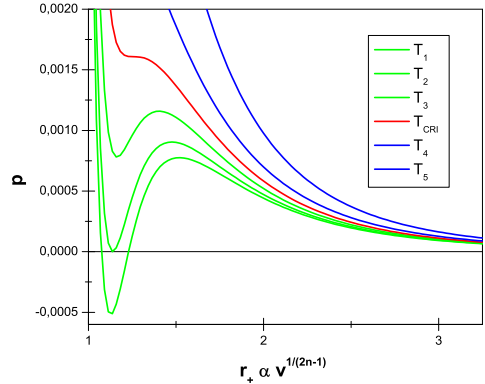}}
\caption{$p-v$ curve for $T_1<T_2<T_3<T_{c}<T_4<T_5$}
\label{p-v}
\end{figure}

In Fig.\ref{p-v} is displayed for $n=3$ and $d=10$ the behavior of the curve $p-v$ defined by equation (\ref{presionPL3}). Although this is an example still this behavior is generic for any values of $n$ and $d$. 

For values of temperature $T>T_{c}$, the second and third factors of Eq.(\ref{presionPL3}) are negligible in comparison with the first one, and therefore the curve approximates the form  $p \cdot v \propto T$ and thus mimicking the behavior of an ideal gas. Conversely, for $T<T_{c}$  as $v$ increases the $p$, which diverges for $v=0$, decreases until reach a local minimum. Next, $p$ increases until reaching a local maximum. Finally $p$ decreases asymptotically until reaching $p=0$. Thus for $T<T_{c}$ ( or for $p<p_{c}$ due that $T \propto p$) the behavior is analogue to the Van der Waals fluid.  

In standard vapor-liquid theory is well established that an increase in pressure must be correlated with a decrease in volume and vice versa. Therefore, see Fig.\ref{p-v}, one can notice the existence of a range of the specific volume, says $]v_{min},v_{max}[$, which must be considered nonphysical due to both pressure and volume increase simultaneously. On the other hand, it can be also noticed that for a single value of the specific pressure, there might exist up to three possible values of specific volume $v$ with one of them always within the nonphysical region. Therefore, for analysis one must only consider the two proper solutions that satisfy either $v_1<v_{min}$ or $v_2>v_{max}$ with $p(v_1)=p(v_2)$. In fluid theory these two solutions are known as the {\it Van der Waals loop} and physically this corresponds to a {\it vapor liquid equilibrium}, highlighting that phase transitions take place. 

\subsubsection{\bf Temperature}
\begin{figure}
\centering
\subfigure[Temperature for $p=1>p_{c}$ .]{\includegraphics[width=75mm]{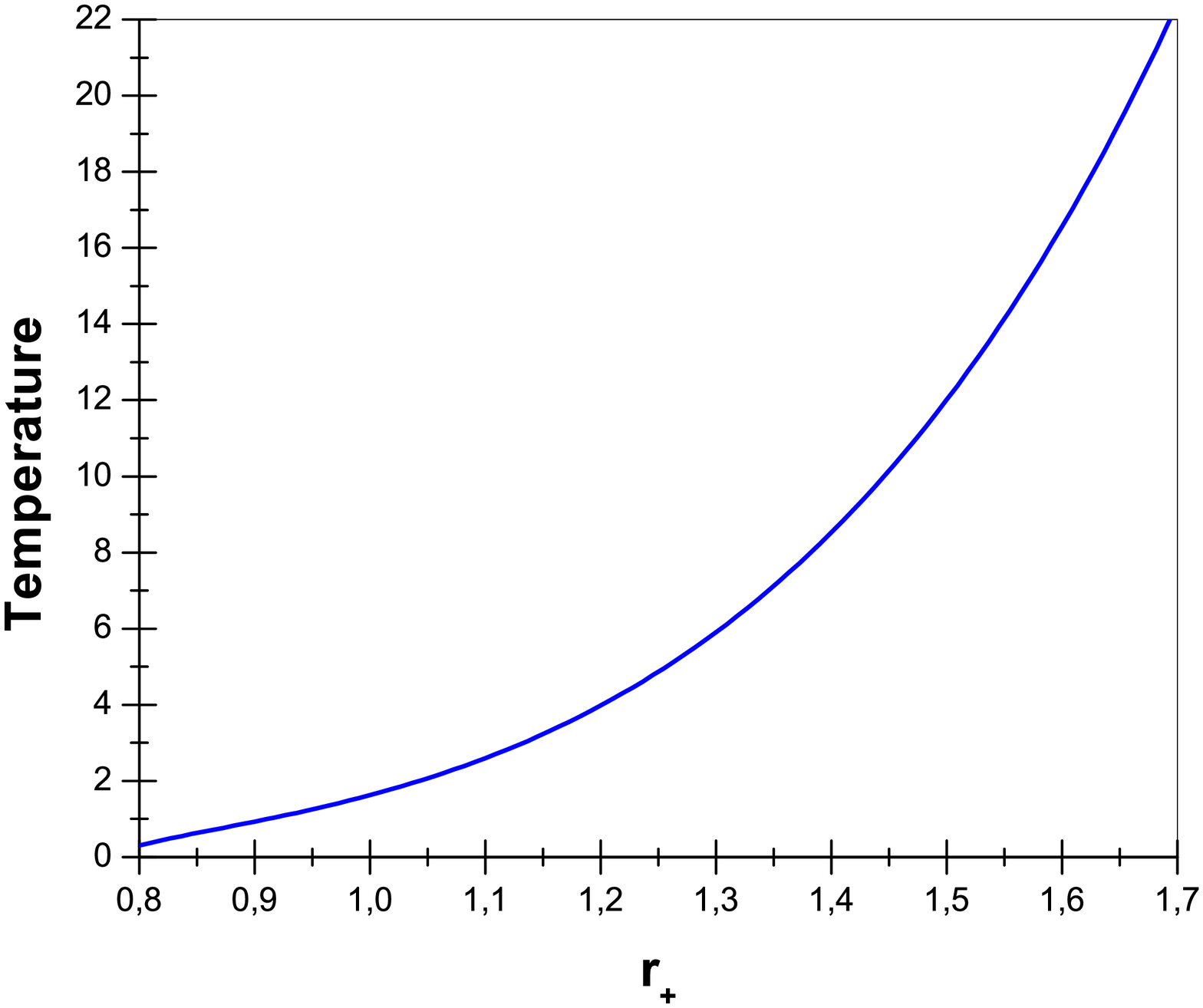}} 
\subfigure[Temperature for $p=p_{c} \approx 0,001608$ ]{\includegraphics[width=75mm]{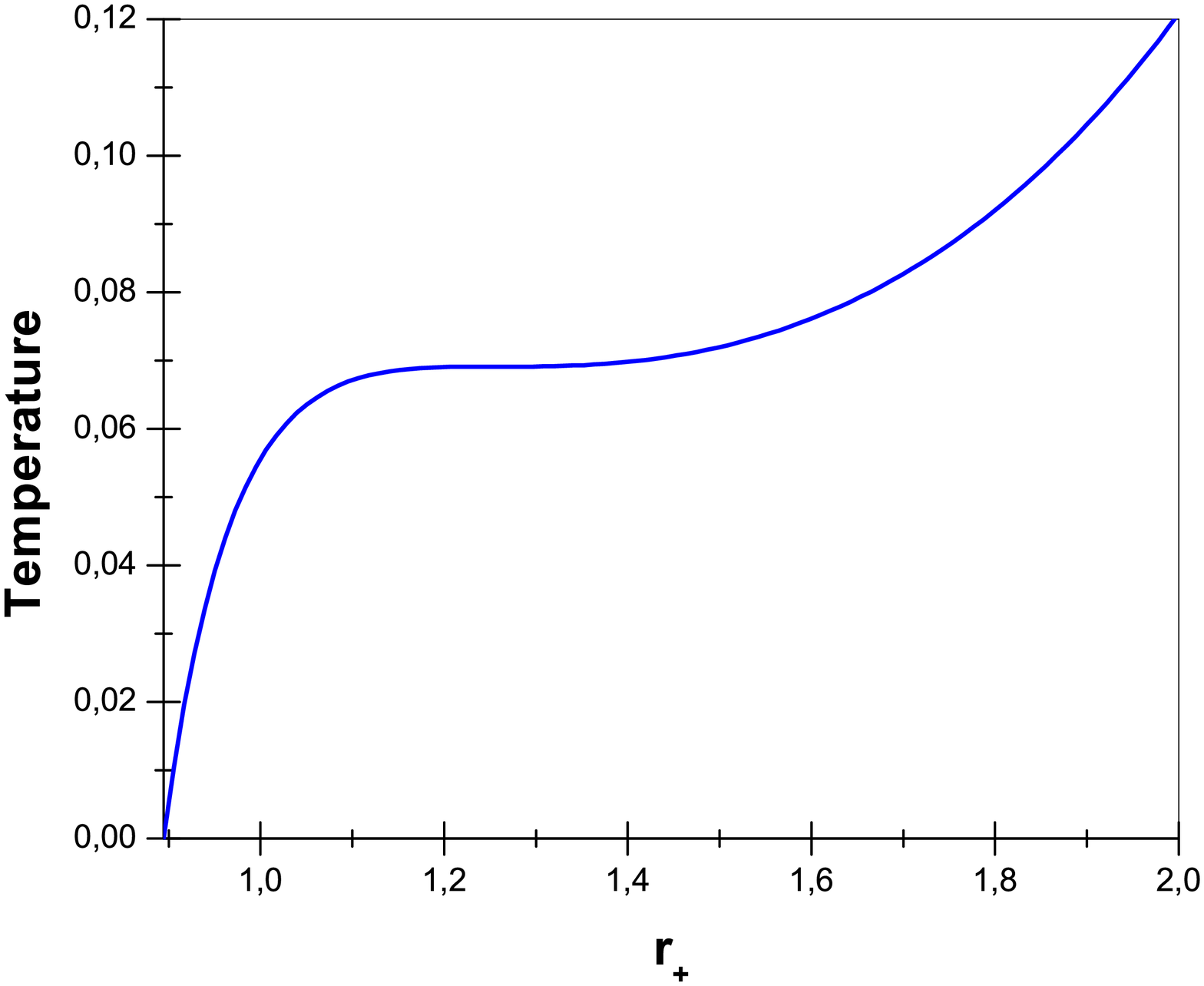}}
\subfigure[Temperature for $p=0,00001<p_{c}$ ]{\includegraphics[width=75mm]{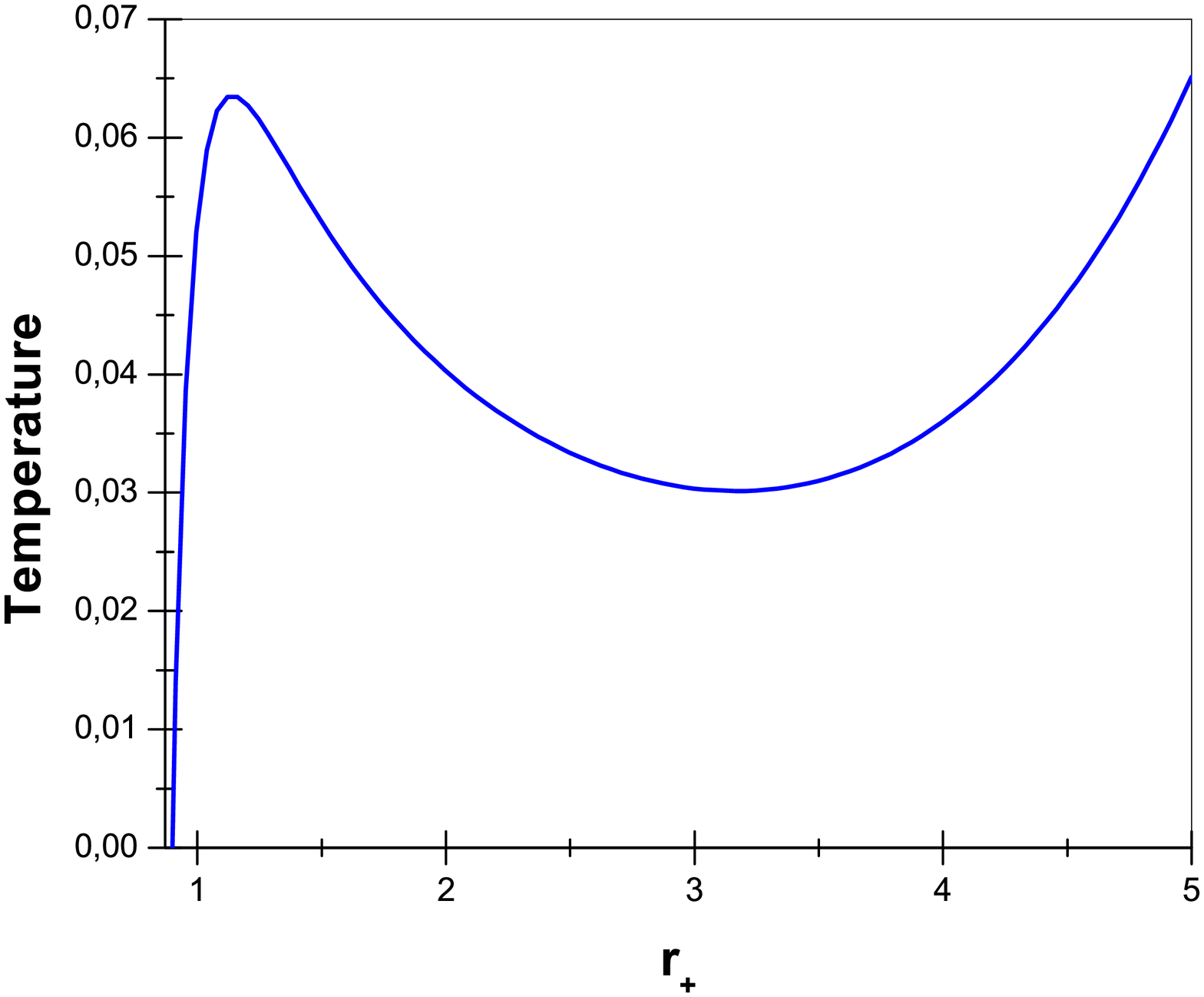}}
\caption{Temperature behavior for $n=3$ and $d=10$ with $Q=1$ .}
\label{FigTemperatura}
\end{figure}

The behavior of the temperature is displayed in figure \ref{FigTemperatura}(a) for $p>p_{c}$, in figure \ref{FigTemperatura}(b) for $p=p_{c}$ and in figure \ref{FigTemperatura}(c) for $p<p_{c}$ . Although for these figures $n=3$ and $d=10$ it is straightforward to show that this enfolds the generic behavior for any $n$ and $d$. One can notice the presence of an extreme black hole case for small $r_+$ where the temperature vanishes. For $p>p_{c}$ the temperature is an increasing function of $r_+$. For $p=p_{c}$ the temperature has one inflexion point at $r_+=r_{infl}$. More relevant for this discussion is the case for $p<p_{c}$, where the fluid is analogous to the Van der Waals, and where the temperature has a local minimum and a local maximum at $r_+=r_{min}$ and $r_+=r_{max}$, respectively.

\subsubsection{\bf Heat Capacity}
\begin{figure}
\centering
\subfigure[Heat capacity for $p=1>p_c$.]{\includegraphics[width=75mm]{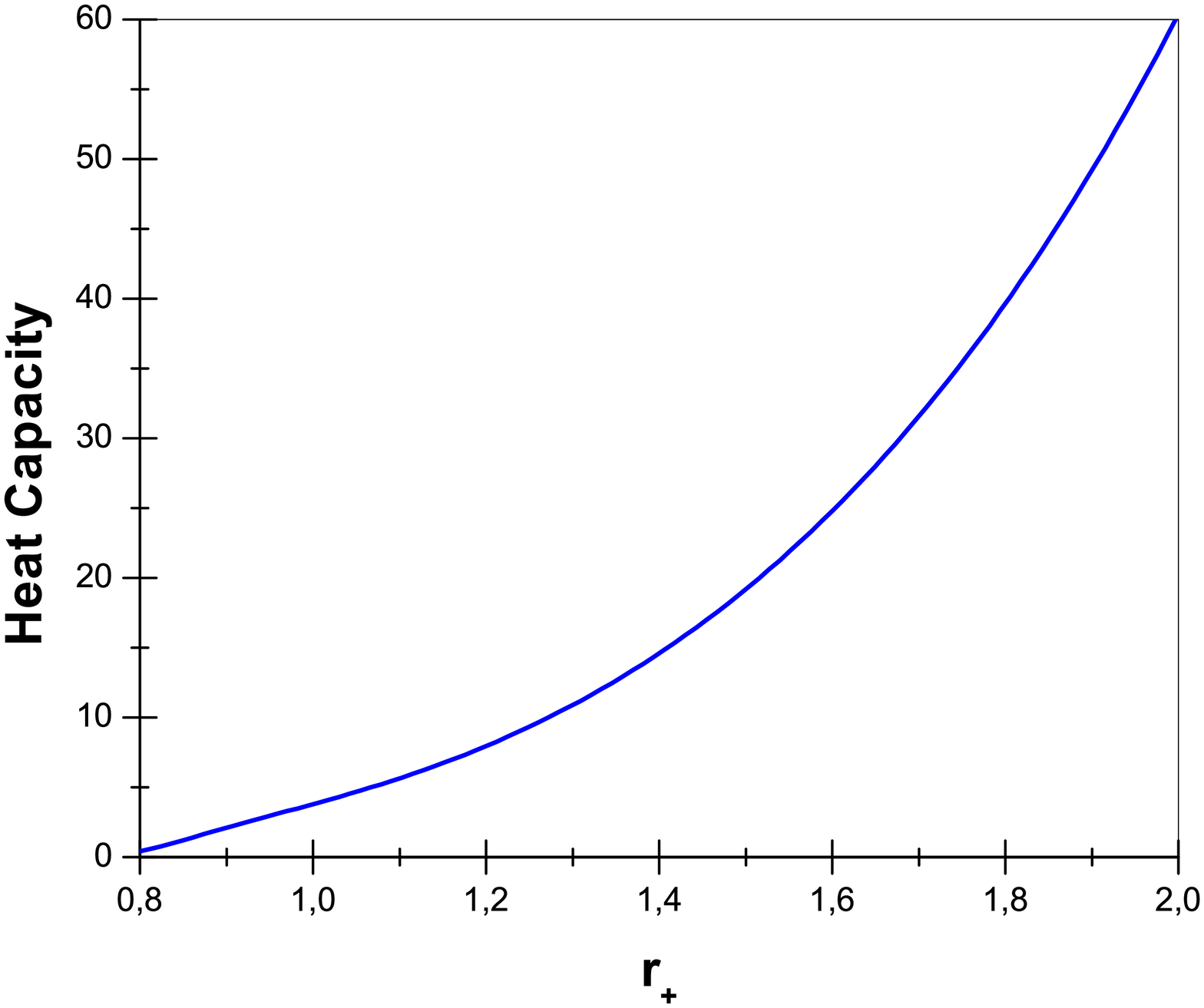}}
\subfigure[Heat capacity for $p=p_c \approx 0,001608 $]{\includegraphics[width=75mm]{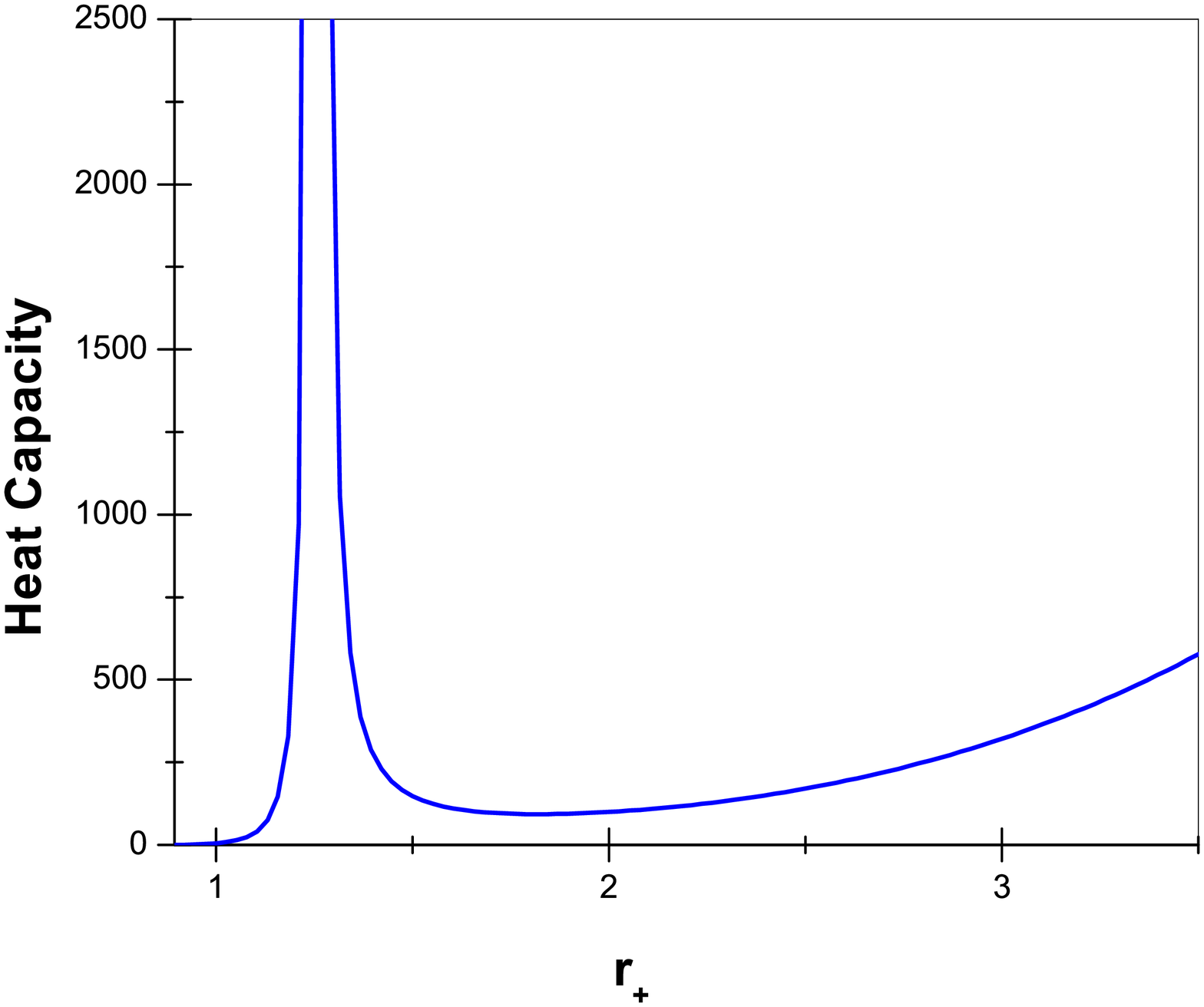}}
\caption{Behavior of Heat Capacity for $n=3$ and $d=10$ with $Q=1$  .}
\label{CalorEspecificoMayoroigual}
\end{figure}

\begin{figure}
\centering
\subfigure[Heat capacity for $r_+ \in (0.896,1.6)$.]{\includegraphics[width=75mm]{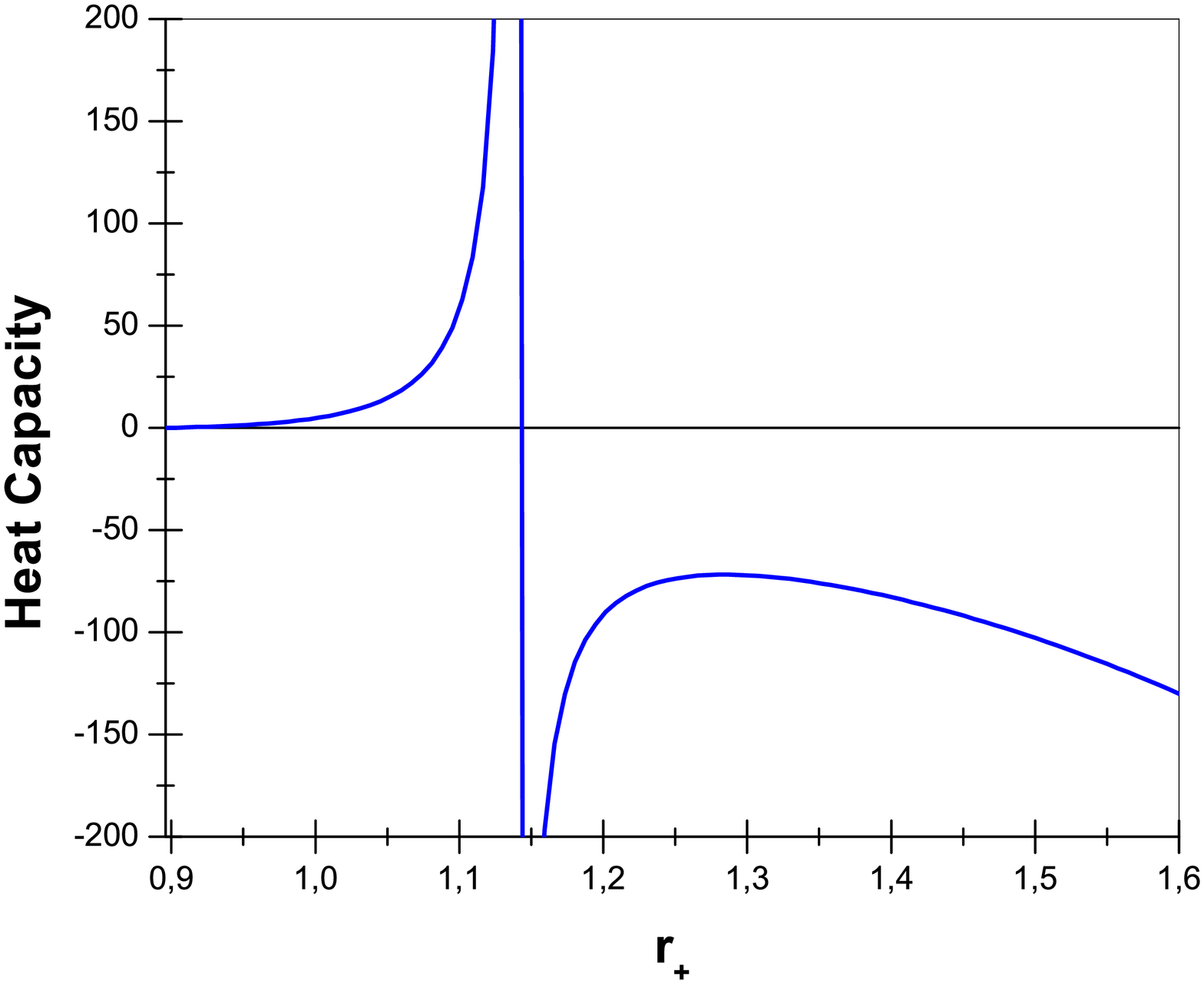}}
\subfigure[Heat capacity for $r_+ \in (2.7,3.5)$]{\includegraphics[width=75mm]{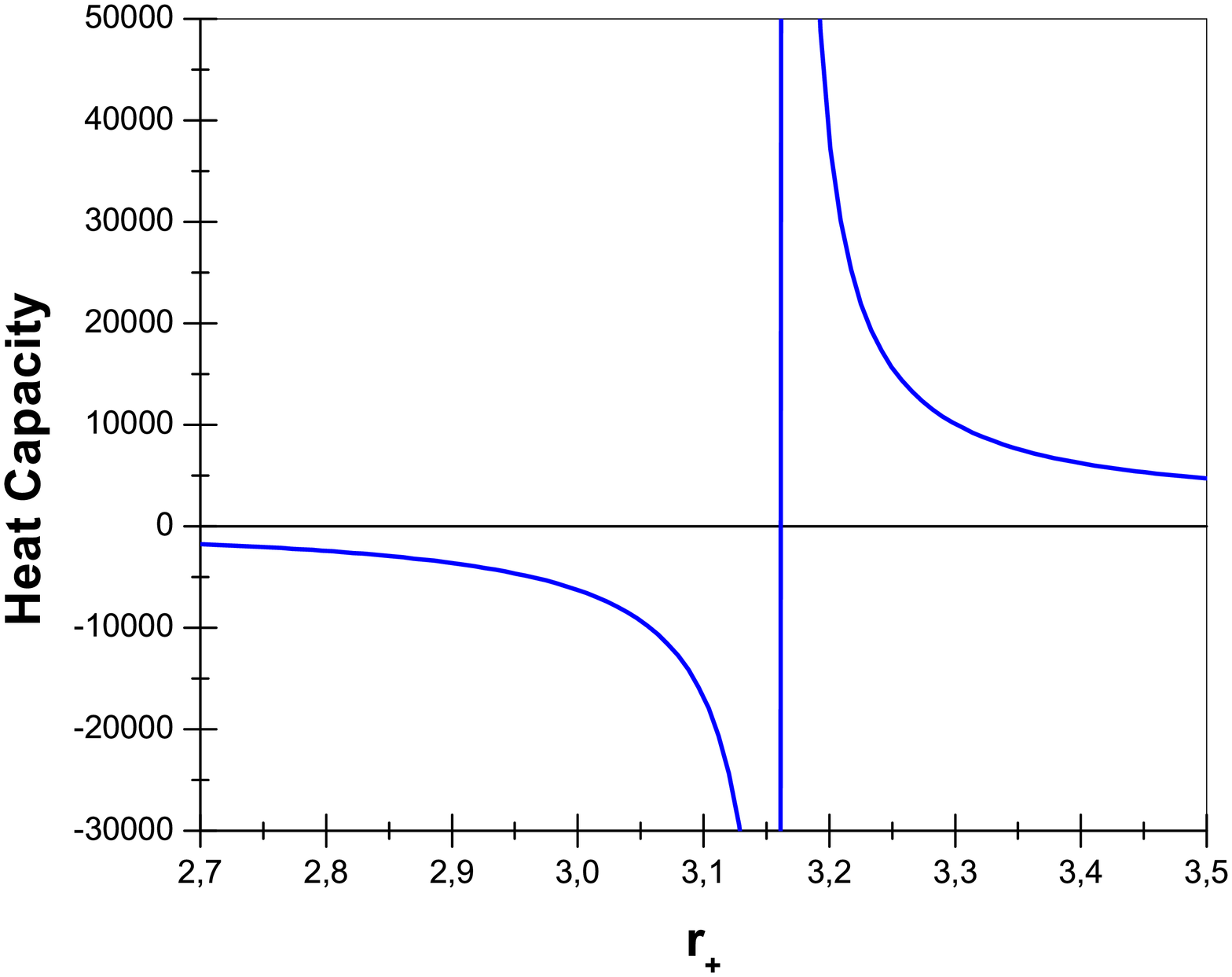}}
\caption{Behavior of Heat Capacity for $n=3$ and $d=10$ with $Q=1$, $p=0,00001<p_c$ .}
\label{CalorEspecificoMenor}
\end{figure}

The heat capacity is displayed in Fig.\ref{CalorEspecificoMayoroigual}(a) for $p>p_c$, in figure \ref{CalorEspecificoMayoroigual}(b) for $p=p_c$ and in figure \ref{CalorEspecificoMenor} for $p<p_c$. As previously, although $n=3$ and $d=10$ it is straightforward to show that Fig.\ref{CalorEspecificoMayoroigual} enfold the generic behavior for any $n$ and $d$. One can notice that 

\begin{itemize}
    \item For $p>p_c$ the heat capacity is a positive increase function of $r_+$, and there is no phase transition, thus black hole is always stable.
    \item For $p=p_c$ small and large black hole coexist at the inflexion point $r_+=r_{infl}$, where the heat capacity $C \to \infty$.
    \item For the $p<p_c$ case, the derivative $(dT/dr_+)$ can vanish for two values of  $r_+=r_{min}$ and $r_+=r_{max}$, as observed in the figure \ref{FigTemperatura}(c). This implies, due to $C= (dS/dr_+)/(dT/dr_+)|_{p,Q}$, that the heat capacity becomes ill-defined at those values of $r_+$. In this way $r_{min}$ and $r_{max}$ define three regions. For $r_+<r_{max}$ one can notice that $C>0$ defining a small stable black hole. Next, there is small unstable ($C<0$) region for $r \in ]r_{max},r_{min}[$. Finally there is a third region $r_+>r_{min}$ where the system is a large stable black hole $(C>0)$. This hints the existence of phase transitions but, by means of the following analysis of the Gibbs free energy, one can check that only the \textit{small stable bh/large stable bh} transition is allowed. 
\end{itemize}

\subsubsection{\bf Gibbs Free energy}
\begin{figure}
\centering
\subfigure[Gibbs for $p=1>p_{c}$ .]{\includegraphics[width=63mm]{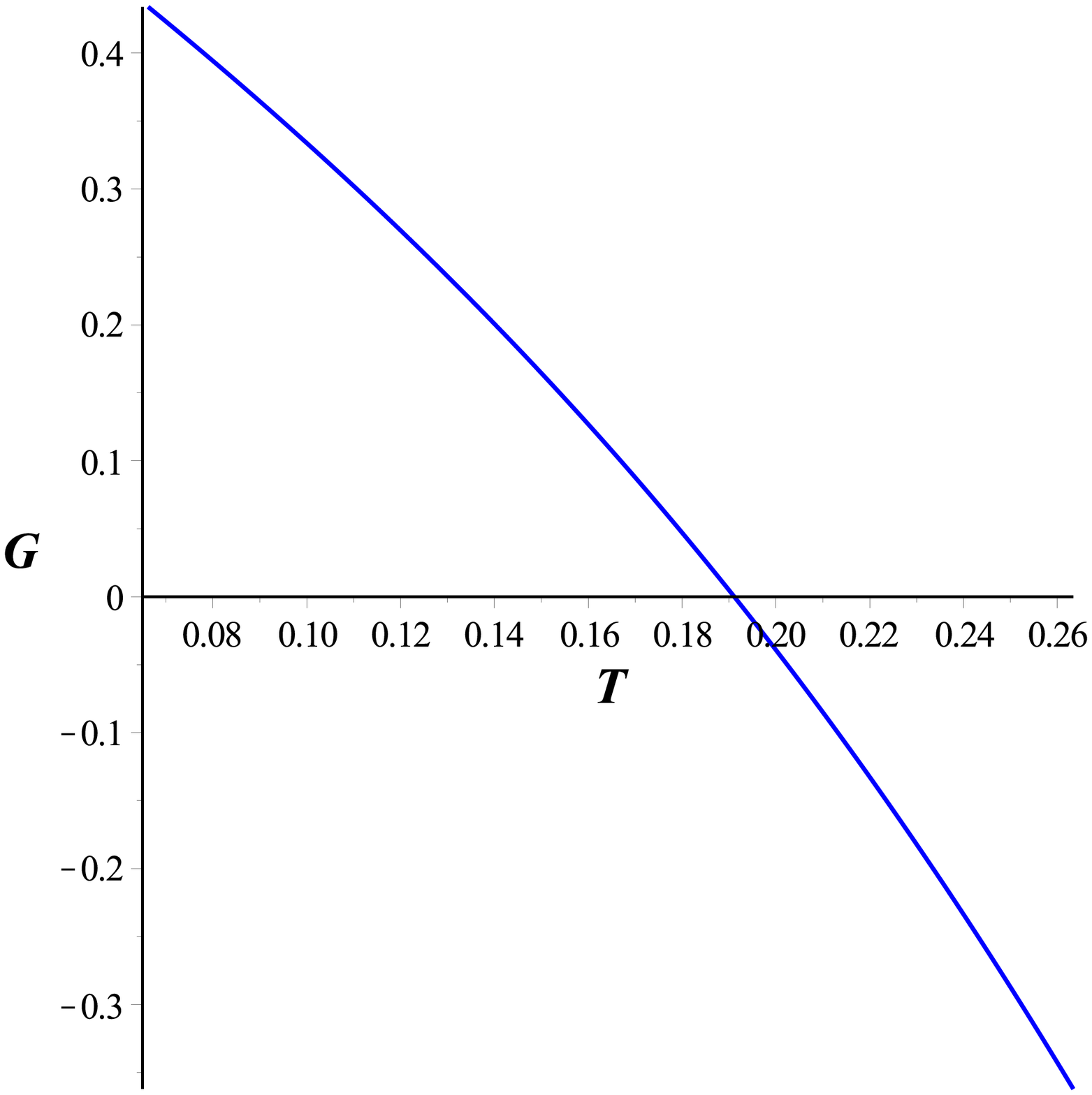}} \hspace{15mm}
\subfigure[ Gibbs for $p=p_{c} \approx 0,001608$ ]{\includegraphics[width=63mm]{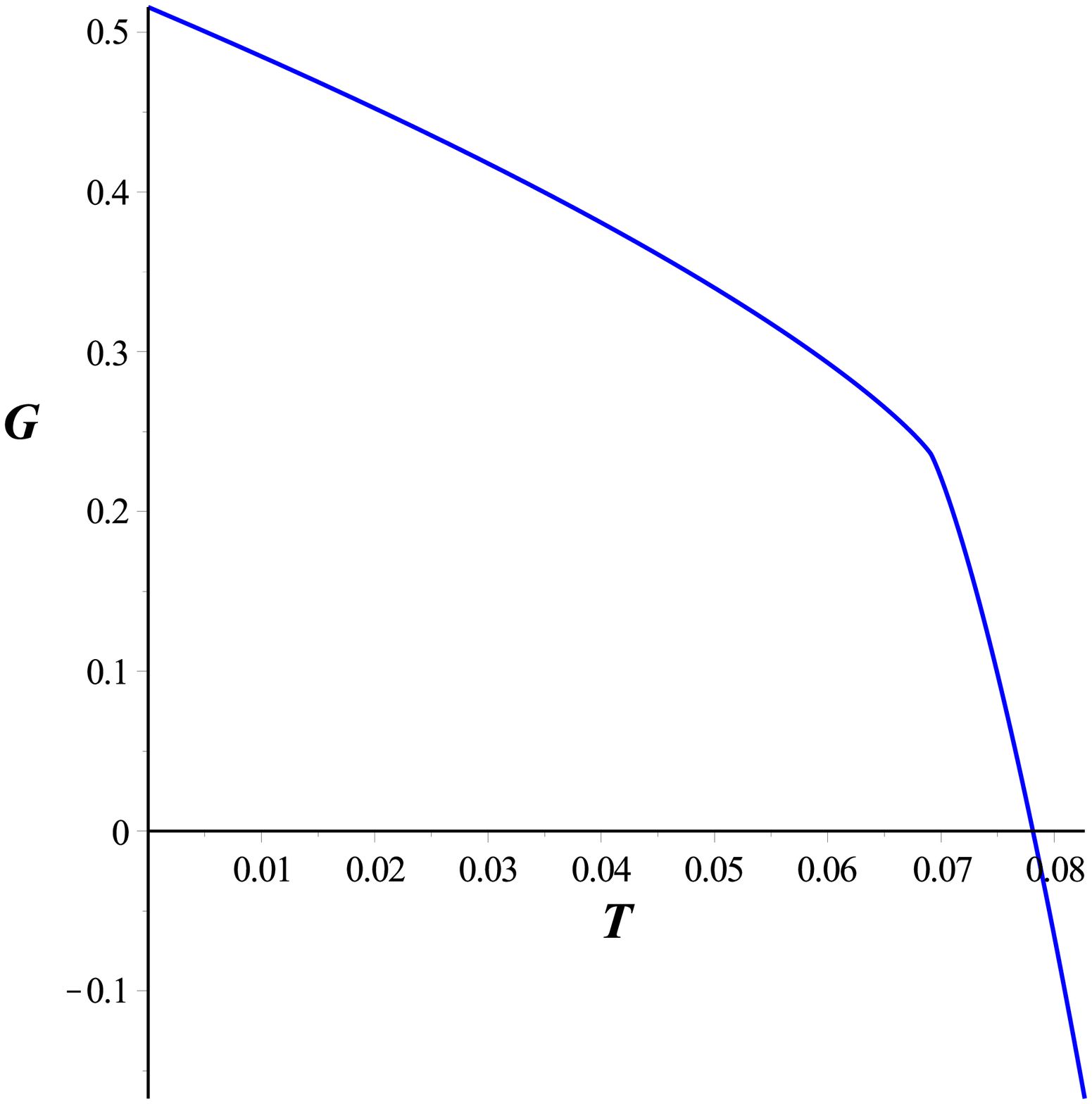}}
\subfigure[Gibbs for $p=0,00001<p_{c}$ ]{\includegraphics[width=63mm]{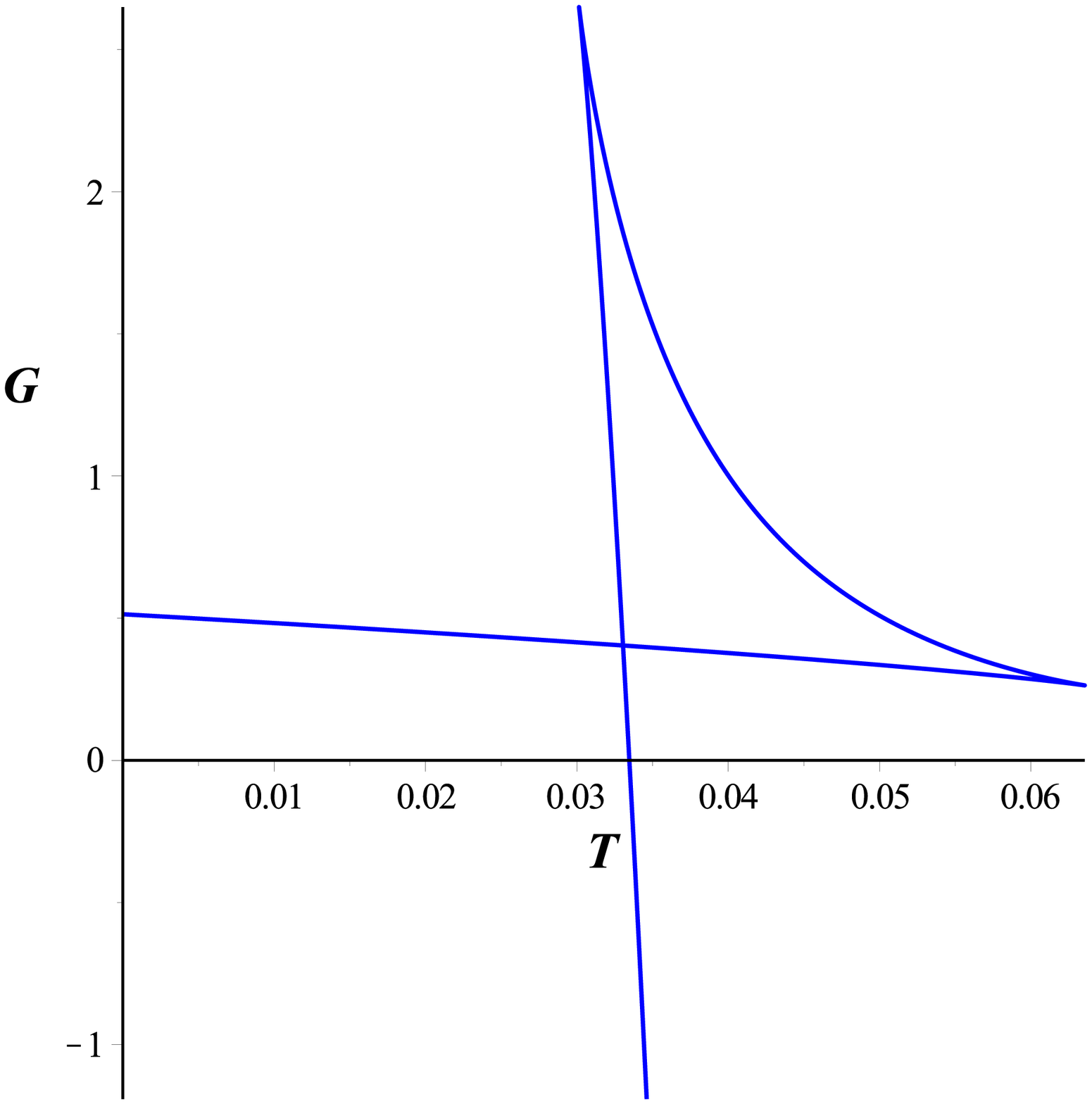}}
\caption{Gibbs free energy (vertical axis) v/s T (horizontal axis) for $n=3$ and $d=10$ with $Q=1$.}
\label{GibbsT}
\end{figure}

\begin{figure}
\centering
{\includegraphics[width=3in]{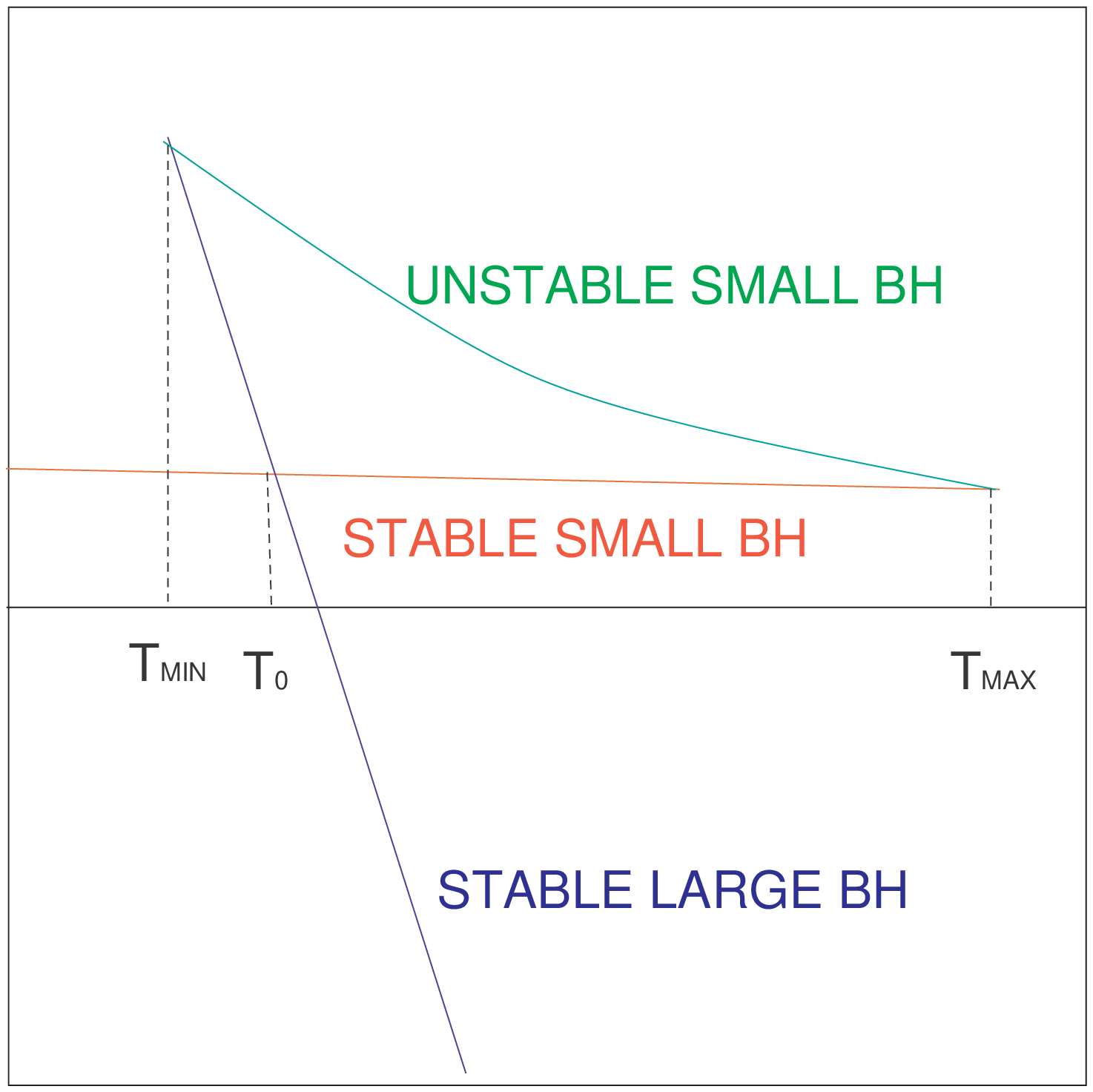}}
\caption{Generic behavior of Gibbs free energy (vertical axis) v/s T (horizontal axis)}
\label{GibbsT1}
\end{figure}

The Gibbs free energy, defined as $G=M-TS$ \cite{Mann1,Hennigar:2018cnh,Czinner:2015eyk,Wang:2019vgz}, is displayed for $n=3$ and $d=10$ in figure \ref{GibbsT}(a) for $p>p_c$, in figure \ref{GibbsT}(b) for $p=p_c$ and in \ref{GibbsT}(c) for $p<p_c$. It is direct to check that this behavior is similar for other values of $n$ and $d$. As well known \cite{RegularBHquimica} the Gibbs free energy describes the global stability of the system. It is worth to recall that the global minimum represents the most likely state, while the preferred state at fixed temperature corresponds to the minimal value of the Gibbs free energy. 
\begin{itemize}
    \item For pressure larger than the critical pressure, the Gibbs free energy is a single valued function.
    \item For $p=p_c$ there is a cusp at $T=T_c$, which coincides with the radius $r_+=r_{infl}$, thus, since at this point $C=T dS/dT=-T (\partial^2G/\partial T^2) \to \infty$, the discontinuity on the second derivative of Gibbs function implies the presence of second order phase transition between small stable/large stable black holes.
    \item The generic behavior of the Gibbs free energy for $p<p_c$ is displayed in figure \ref{GibbsT1}. We see three possibles black hole states: small stable, small unstable and large stable. The preferred state is such that the Gibbs free energy has the minimum value. Thus, for $]T_0,T_{max}]$ the preferred state correspond to stable large black hole. However, for $T<T_0$ the preferred state corresponds to the small stable black hole. Thus, at $T=T_0$ there is a first order phase transition between large/small stable black hole.
\end{itemize}

\subsubsection{\bf Behavior near critical points}
Let's define the following dimensionless variables 
\begin{equation} \label{omega}
    \omega =\left ( \frac{V}{V_c} \right)^{2n-1}-1=\left ( \frac{v}{v_c} \right)^{d-1}-1
\end{equation}
and 
\begin{equation} \label{t}
    t = \frac{T}{T_c}-1,
\end{equation}
to analyze the behavior nearby the critical points. It is direct to check that near the critical points, \textit{i.e.} $V \to V_c$ (or $v \to v_c$) and $T \to T_c$, the variables $\omega \to 0$, and $t \to 0$, respectively. The pressure in Eq.(\ref{presionPL3}) is displayed in table \ref{tabla2} at $O(t\omega^2, \omega^4)$. The truncation will be justified below . 
\begin{table*} 
\caption{Behavior of pressure near critical points .}
\begin{tabular*}{\textwidth}{@{\extracolsep{\fill}}llc@{}}
\hline
\multicolumn{1}{l}{$n$   } & \multicolumn{1}{l}{$d$ } & \multicolumn{1}{c}{$p$} \\
 \hline
$1$ & $4$  & $p \approx 1 + 8/3 t - 8/9t \omega -4/81 \omega^3$    \\
\hline 
$1$ & $5$  & $p \approx 1 + 12/5t - 3/5t \omega -1/32 \omega^3$  \\
\hline
$1$ & $6$  & $p \approx 1 + 16/7t - 16/35t \omega -8/375 \omega^3$   \\
\hline
$3$ & $8$  & $p \approx 1 + 72/7t - 72/49t \omega -12/8575 \omega^3$   \\
\hline
$3$ & $9$  & $p \approx 1 + 28/3t - 7/6t \omega -7/6400 \omega^3$   \\
\hline
$3$ & $10$  & $p \approx 1 + 96/11t - 32/33t \omega -16/18225 \omega^3$   \\
\hline
$5$ & $12$  & $p \approx 1 + 200/11t - 200/121t \omega -100/323433 \omega^3$    \\
\hline
\end{tabular*} \label{tabla2}
\end{table*}

In general it is possible to approximate the pressure as
\begin{equation} \label{presiongenerica}
    p \approx 1 + \frac{n}{Z}t - \frac{n t}{(d-1)Z} \omega - \frac{(2d-2n-3)n}{6Z(d-1)^3(2n-1)^2} \omega^3,
\end{equation}
whereas 
\begin{equation}
    dP= -\frac{P_c n}{Z (d-1)} \left ( t+ \frac{(2d-2n-3)}{2(d-1)^2(2n-1)^2} \omega^2 \right ) d \omega.
\end{equation}
To compute the \textit{ critical exponents} one can follow \cite{Mann2}. 
\begin{itemize}
    \item The {\it $\alpha$ exponent} describes the behavior of the heat capacity at constant volume defined as
    \begin{equation}
        C_v = T \frac{\partial S}{\partial T} \propto |t|^{-\alpha}.
    \end{equation}
    In the case at hand, since entropy and volume are both functions of horizon radius, see Eqs.(\ref{entropia1},\ref{volumen11}), then a constant volume implies a constant entropy as well. Therefore it is satisfied that $C_v=0$ which implies that there is no dependence on $|t|$. In turn this implies that
    \begin{equation}
        \alpha=0.
    \end{equation}
  \item The \textit{exponent} $\beta$ describes the behavior of the \textit{order parameter} defined as
  \begin{equation} \label{eta}
      \eta = V_l-V_s \propto |t|^\beta.
  \end{equation}
  To compute this order parameter one can use the \textit{Maxwell’s area law}
\begin{equation} \label{integral1}
    \oint VdP=0,
\end{equation}
where, the volume (\ref{omega}) is approximated such that $VdP$ is truncated to $O(t\omega^3, \omega^5)$, in other words, 
\begin{equation} \label{aproxVolumen}
    V \approx V_c \left (  1 + \frac{1}{2n-1} \omega   \right ).
\end{equation}
Therefore, closed integral (\ref{integral1}) becomes
\begin{equation} \label{integral2}
  V_c  \oint  dP + \frac{V_c}{2n-1} \oint \omega dP =0.
\end{equation}
The second integral of the left side of (\ref{integral2}) yields
\begin{equation}
 \int^{\omega_s}_{\omega_l}\omega dP= \int^{\omega_s}_{\omega_l} \omega   \left ( t+ \frac{(2d-2n-3)}{2(d-1)^2(2n-1)^2} \omega^2 \right ) d \omega =0.
\end{equation}
which has the non trivial solution given by
\begin{equation} \label{condicion}
    \omega_s = -\omega_l.
\end{equation}
On the other hand, the first integral of the left side of (\ref{integral2}) yields
\begin{align}
    &\int^{\omega_s}_{\omega_l}dP = 0 \nonumber \\
&1 + \frac{n}{Z}t - \frac{n t}{(d-1)Z} \omega_l - \frac{(2d-2n-3)n}{6Z(d-1)^3(2n-1)^2} \omega^3_l = \nonumber \\
&1 + \frac{n}{Z}t - \frac{n t}{(d-1)Z} \omega_s - \frac{(2d-2n-3)n}{6Z(d-1)^3(2n-1)^2} \omega^3_s,
\end{align}  
This, by condition (\ref{condicion}), yields
\begin{equation} \label{omegal}
    \omega_l=(d-1)(2n-1) \sqrt{- \frac{6}{2d-2n-3}t},
\end{equation}
 with $t<0$. Replacing Eqs. (\ref{aproxVolumen}) and (\ref{omegal}) into Eq.(\ref{eta}) one can obtain 
 \begin{equation}
     \eta = \frac{2V_c}{2n-1} \omega_l = 2(d-1)V_c\sqrt{- \frac{6}{2d-2n-3}t},
 \end{equation}
 Comparing this result with Eq.(\ref{eta}) one can uncover that 
 \begin{equation}
     \beta = \frac{1}{2}.
 \end{equation}
\item Now one can compute the \textit{exponent} $\gamma$ which describes the behavior  under \textit{isothermal compressibility}, $\kappa_T$, defined by
\begin{equation}
\kappa_T= - \frac{1}{V}\frac{\partial V}{\partial P} \bigg |_T \propto |t|^{-\gamma}.
\end{equation}

By using Eqs.(\ref{presiongenerica},\ref{aproxVolumen}) one can prove that
\begin{equation}
    \frac{\partial P}{\partial V}=P_c \frac{\partial p}{\partial \omega} \frac{\partial \omega}{\partial V} \propto -(2n-1)\frac{P_c}{V_c} \frac{n}{(d-1)Z}t,
\end{equation}
and therefore
\begin{equation}
    \kappa_T \propto \frac{(d-1) Z}{P_c n(2n-1)t},
\end{equation}
from which is direct to read that
\begin{equation}
    \gamma=1.
\end{equation}

\item Finally, one can compute \textit{exponent} $\delta$ which describes the behavior on the critical isotherm $T=T_c$, and therefore for $t = 0$. In this case this is defined by
\begin{equation}
    |P-P_c| \propto |V-V_c|^\delta.
\end{equation}
From Eq.(\ref{presiongenerica}), at $t=0$, it is possible to notice that
\begin{equation} \label{presiongenerica1}
    p -1 \approx - \frac{(2d-2n-3)n}{6Z(d-1)^3(2n-1)^2} \omega^3.
\end{equation}
Using the approximation of Eq.(\ref{aproxVolumen}) one can show that
\begin{equation}
    \frac{P-P_c}{P_c} \approx - \frac{(2d-2n-3)n}{6Z(d-1)^3(2n-1)^2} \left ( (2n-1) \frac{V-V_c}{V_c}   \right )^3,
\end{equation}
and therefore
\begin{equation}
    \delta=3. 
\end{equation}
\end{itemize}

This critical exponents just computed are similar to those of Van der Waals gas. Although the presence of extra dimensions and the value of $n$ modify the value of the compressibility factor respect to the well known value $Z=3/8$, they do not affect the value of the critical exponents, and thus, the behavior is still similar to that of Van der Waals fluid near the critical exponents.

\section{Conclusion and discussion} 
In this article it has been analyzed the thermodynamics of the Pure Lovelock solutions in $d$ dimensions, in an extended phase space including the introduction of pressure and volume as dual thermodynamic variables and the mass parameter standing for the enthalphy of the system. A linear relation between the cosmological constant and the thermodynamics pressure, valid for all value of $n$ (odd) and $d$, has been established. The thermodynamic volume obtained corresponds to a generalization of the black hole \textit{volume}. The Smarr formula for Pure Lovelock gravity was constructed as well.

In addition, it was shown that the variation of parameters are constrained to follow a first law of thermodynamics in the extended phase space. The temperature is defined by surface gravity, as usual, the entropy coincides with the value computed {\it a la Wald} \cite{Cai:2006pq, milko1} and the electric potential matches the usual known definition. 

The equation state for the uncharged and charged cases was obtained. For the uncharged case, a Hawking-Page-Like phase transitions between thermal radiation and large stable black hole has been found. For the charged case, it was found that the compressibility factor, $Z$, is a generic function of $d$ and $n$ given by equation (\ref{factorcompresibilidad}). Remarkably, $Z<1$ estrictly and therefore the behavior always corresponds to a real gas. 

It was found the existence of a critical temperature, $T_c$, where phase transitions occur. The mapping of the $p-v$ curves indicate that for Pure Lovelock gravity the behavior is similar to an ideal gas for $T>T_c$. For $T<T_c$ the behavior is analogous to a Van der Walls fluid. Furthermore, there are a first order phase transition between small stable/large stable black hole, which are analogous to liquid/gas phase transitions.

Finally, we have computed the critical exponents whose values are similar to those of the Van der Waals gas. Furthermore, it is possible to write the pressure as a generic function of $n$ and $d$ given by equation (\ref{presiongenerica}). Thus, the presence of extra dimensions and the value of $n$ does not modify the known values of the critical points, therefore the general behavior can be considered similar to a  Van der Waals fluid near the critical exponents.

\section*{Acknowledgments}
This work was partially funded by grant DI-08-19 UNAB.

\bibliography{mybibHeatEngine}

\end{document}